\documentclass[sigconf]{acmart}
\usepackage{amsmath, array, graphicx}
\usepackage{multirow}
\usepackage{enumitem}
\usepackage{booktabs}
\usepackage{tabularx}
\usepackage{subcaption}
\usepackage{balance}
\usepackage{makecell}
\usepackage[nolist,nohyperlinks]{acronym}
\usepackage[]{siunitx}
\usepackage{svg}
\usepackage{colortbl}

\definecolor{ballblue}{rgb}{0.13, 0.67, 0.8}
\definecolor{rhodamine}{rgb}{0.8, 0.2, 0.8}

\definecolor{softblue}{rgb}{33, 171, 204}
\definecolor{strongblue}{rgb}{204, 51, 204}

\newcommand{\mycaption}[1]{\caption{\normalfont{#1}}}
\newcommand{\myparagraph}[1]{\vspace{0.5\baselineskip}\noindent{\textbf{#1}}.~}
\newcommand{\boldheadpara}[1]{\vspace{0.25\baselineskip}\noindent{\textbf{#1}}}

\acrodef{EX}[EX]{\emph{Example}}
\acrodef{IR}{Information Retrieval}
\acrodef{NLP}{Natural Language Processing}
\acrodef{DL21}{Deep Learning Track of TREC 2021}
\acrodef{DL22}{Deep Learning Track of TREC 2022}
\acrodef{LLM}{Large Language Model}
\acrodef{MAE}{Mean Absolute Error}
\acrodef{NIST}{National Institute of Standards and Technology}
\acrodef{AA}{Active Agreement}
\acrodef{AD}{Active Disagreement}
\acrodef{PA}{Passive Agreement}
\acrodef{PD}{Passive Disagreement}
\acrodef{MA}{Mixed Agreement}
\acrodef{MD}{Mixed Disagreement}

\copyrightyear{2025}
\acmYear{2025}
\setcopyright{cc}
\setcctype{by}
\acmConference[ICTIR '25]{Proceedings of the 2025 International ACM SIGIR Conference on Innovative Concepts and Theories in Information Retrieval (ICTIR)}{July 18, 2025}{Padua, Italy}
\acmBooktitle{Proceedings of the 2025 International ACM SIGIR Conference on Innovative Concepts and Theories in Information Retrieval (ICTIR) (ICTIR '25), July 18, 2025, Padua, Italy}\acmDOI{10.1145/3731120.3744608}
\acmISBN{979-8-4007-1861-8/2025/07}

\begin{document}

\title{Demographically-Inspired Query Variants Using an LLM}

\author{Marwah Alaofi}
\orcid{https://orcid.org/0000-0002-0008-8650}
\affiliation{
    \institution{RMIT University}
    \city{Melbourne}
    \country{Australia}
}
\email{marwah.alaofi@student.rmit.edu.au}

\author{Nicola Ferro}
\orcid{https://orcid.org/0000-0001-9219-6239}
\affiliation{
\institution{University of Padua}
\city{Padova}
\country{Italy}
}
\email{ferro@dei.unipd.it}

\author{Paul Thomas}
\orcid{https://orcid.org/0000-0003-2425-3136}
\affiliation{
\institution{Microsoft}
\city{Adelaide}
\country{Australia}
}
\email{pathom@microsoft.com}

\author{Falk Scholer}
\orcid{https://orcid.org/0000-0001-9094-0810}
\affiliation{
\institution{RMIT University}
\city{Melbourne}
\country{Australia}
}
\email{falk.scholer@rmit.edu.au}

\author{Mark Sanderson}
\orcid{https://orcid.org/0000-0003-0487-9609}
\affiliation{
\institution{RMIT University}
\city{Melbourne}
\country{Australia}
}
\email{mark.sanderson@rmit.edu.au}

\begin{abstract}
This study proposes a method to diversify queries in existing test collections to reflect some of the diversity of search engine users, aligning with an earlier vision of an `ideal' test collection. A \ac{LLM} is used to create query variants: alternative
queries that have the same meaning as the original. These variants represent user profiles characterised by different properties, such as language and domain proficiency, which are known in the \ac{IR} literature to influence query formulation.

The \ac{LLM}'s ability to generate query variants that align with user profiles is empirically validated, and the variants’ utility is further explored for \ac{IR} system evaluation. Results demonstrate that the variants impact how systems are ranked and show that user profiles experience significantly different levels of system effectiveness. This method enables an alternative perspective on system evaluation where we can observe \textit{both} the impact of user profiles on system rankings and how system performance varies across users.
\end{abstract}



\maketitle

\section{Introduction}

The `ideal' test collections envisioned by Sp{\"a}rck-Jones and Van Rijsbergen~\cite{Sparckjones1975Ideal} called for diversity in both documents \emph{and} `requests' (now known as queries) by varying content, type, source, and origin. This vision reflects \citeauthor{Cleverdon1961Application}'s early emphasis on including `representative' questions in evaluation. Although the creation of test collections -- such as those developed in TREC -- was largely inspired by this vision, it has not been fully realised. Most focus has been on document pooling and consistency of relevance judgments; far less attention has been given to the diversity of queries.

Past work has shown that factors like modality \cite{Church2007Mobile, Harvey2017Fragmented}, domain expertise \cite{Monchaux2015expert,White2009Expert}, age \cite{Torres2010children, BILAL2018Children}, and language proficiency \cite{Chu2015non-native} influence query formulation, and consequently affect the accuracy of search results and user satisfaction. 
However, both capturing this variation and understanding its utility in test collection remain limited. Most system evaluation is based on so-called generic queries; how systems perform in other contexts, e.g., mobile users, or those with limited domain expertise, remains under-explored. This raises the question: how would the effectiveness of \ac{IR} systems differ under a broader spectrum of queries? This calls for innovative approaches to develop cost-effective, comprehensive evaluation environments.

This work seeks to realise the original vision of the `ideal' test collections by reintroducing the often-overlooked element of user diversity -- explicitly controlling for variables that influence query formulation. The work describes an approach to expand a one query representation of information needs to better represent a diverse range of search engine users. It introduces a simulation of query \textit{variants}: alternative queries that have the same meaning as an original (seed) query. The simulated incorporation of user properties is achieved through three \ac{LLM}-based transformation methods applied to seed queries. These methods integrate profiles for specific personas (e.g., \textit{Anne} the retired nurse), user groups (e.g., \textit{Voice} search users), and single textual transformations (e.g., \textit{paraphrasing}). Table~\ref{tbl-example-queries} shows examples of variants from the three methods, along with their effectiveness scores on two \ac{IR} systems. 

\begin{figure*}
    \centering
    \setlength{\abovecaptionskip}{1.2pt}
    \setlength{\belowcaptionskip}{1.2pt}
    \includegraphics[width=0.9\textwidth]{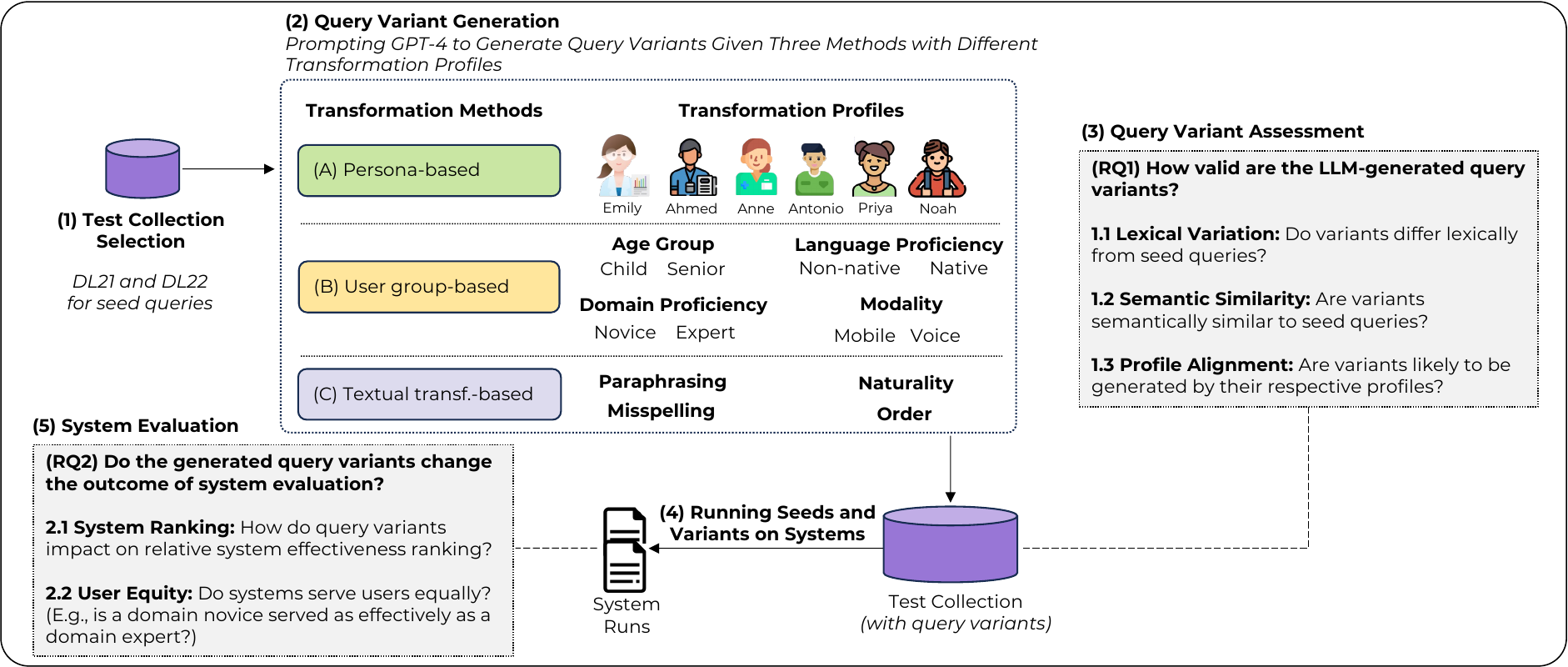}
    \mycaption{An overview of the experiment pipeline and the research questions.}
    \label{fig-overview}
\end{figure*}

The work aims to \textit{validate} \ac{LLM}-generated query variants and explore their \textit{retrieval impact}. A variant is deemed \textit{valid} if it is formulated differently yet shares the same meaning as the seed query, and if it reasonably manifests its profile. The study explores the impact of variants on the effectiveness of IR systems, and offers new insights regarding their utility in system evaluation. An overview of the experimental design is given in Figure \ref{fig-overview}. We explore the following research questions:

\boldheadpara{RQ1 Are \ac{LLM}-generated query variants valid?}
\begin{itemize}[topsep=0pt, partopsep=0pt, itemsep=0pt, parsep=0pt, left=1.5em]
    \item [\textbf{1.1}] Lexical Variation: Do variants differ lexically from seed queries?
    \item [\textbf{1.2}] Semantic Similarity: Are variants semantically similar to seed queries?
    \item [\textbf{1.3}] Profile Alignment: Are variants likely to be generated by their respective profiles?
\end{itemize}
        
\boldheadpara{RQ2 Do the generated query variants change the outcome of system evaluation?}
\begin{itemize}[topsep=0pt, partopsep=0pt, itemsep=0pt, parsep=0pt, left=1.5em]
    \item [\textbf{2.1}] System Ranking: How do variants impact on relative system effectiveness rankings? 
    \item [\textbf{2.2}] User Equity: Do systems serve users equally? (E.g., is a mobile user served as effectively as a voice search user?)
\end{itemize}

We make the following contributions in understanding the utility of an \ac{LLM} in simulating queries and creating relevance labels: 
\footnote{Data is available at: \url{https://github.com/MarwahAlaofi/demo-qv}}

\begin{itemize}
    \item [\textbf{C1}] An adaptable evaluation pipeline that, based on search engine user properties, generates query variants for fine-grained system evaluations. This assists in highlighting problems faced by certain demographics, and identifying where the system performs effectively or inadequately.
    \item [\textbf{C2}] A publicly available dataset of approximately 7K query variants for the \ac{DL21} and \ac{DL22}, with human assessments covering a 10\% sample of each profile, to support studying similarity and profile alignment.
    \item [\textbf{C3}] An evaluation of GPT-4o prompted relevance labels for \ac{DL21} and \ac{DL22}. The labels -- 146K query-passage pairs -- are made available publicly, with ~13K labels human-judged by the \ac{NIST}, enabling further research into \ac{LLM} relevance labels.
\end{itemize}

\begin{table}
\centering
\small
\mycaption{An example seed query from DL22 and its LLM-generated query variants and their NDCG@10 scores using two \ac{IR} systems.}
\label{tbl-example-queries}
\begin{tabularx}{\columnwidth}{m{1.8cm} X m{0.7cm} m{0.7cm}}
\toprule
Profile & Variant & BM25 & ANCE \\
\midrule
\textbf{Seed} & \textbf{What to do if antibiotics cause nausea} & \textbf{0.23} & \textbf{0.35} \\
\midrule

\makecell[tl]{Emily\\\textit{the scientist}} 
& \parbox[t]{\hsize}{\raggedright \textit{\textbf{Coping strategies} for nausea resulting from antibiotic use}} 
& 0.00 & 0.51 \\

\makecell[tl]{Noah\\\textit{the curious child}} 
& \parbox[t]{\hsize}{
    \vspace*{-1.2\baselineskip}%
    \raggedright \textit{How can I feel better if medicine for sickness makes me \textbf{throw up}?}
} 
& \parbox[t]{0.7cm}{\vspace*{-1.2\baselineskip}0.19} 
& \parbox[t]{0.7cm}{\vspace*{-1.2\baselineskip}0.10} \\

\midrule
Non-native & \textit{What should one do when antibiotics \textbf{make feel sick}} & 0.07 & 0.13 \\
Voice & \textit{\textbf{Please find solutions} for dealing with nausea caused by antibiotic medication} & 0.26 & 0.63 \\
\midrule
Paraphrasing & \textit{What to do when antibiotics make you \textbf{queasy}} & 0.35 & 0.50 \\
Naturality & \textit{\textbf{Treatment} for nausea from antibiotics} & 0.15 & 0.47 \\
\bottomrule
\end{tabularx}
\end{table}

\section{Related Work}
We examine two aspects of past work.

\subsection{Users and Search Queries}
The relationship between user characteristics, context, and search behaviour has been widely studied. For example, mobile search queries are often shorter than desktop ones, likely due to limited input options on mobile devices \cite{Church2007Mobile}. The accuracy of these queries is also compromised by the fragmented attention of users, resulting in fewer hits and lower performance~\cite{Harvey2017Fragmented}. On the other hand, voice queries, which are increasingly prevalent, are generally longer and more closely resemble natural spoken language \cite{Guy2016Voice}.

Language proficiency plays a crucial role in the success of online search queries. As highlighted in a qualitative study by \citet{Chu2015non-native}, non-native English speakers faced challenges in formulating queries for English content. Chinese participants had difficulty expressing specialised or academic terms and opted for simpler keywords to reduce inaccuracies. Hungarian participants exhibited a similar pattern, preferring shorter queries in English. 
Strategies such as translating native language queries into English, and beginning with simple keywords and progressively refining them by browsing search results, are used, which affect the accuracy of searches and increase the time taken to find relevant information.

Similarly, research on domain expertise has identified distinct differences in search behaviours between experts and novices. 
For example, \citet{White2009Expert} demonstrated that experts typically formulate longer queries and employ more technical vocabulary from domain-specific lexicons, 50\% more likely than novices. More recently, \citet{Monchaux2015expert} showed that psychology experts outperformed novices in finding correct answers, especially in complex searches. This success was mainly attributed to their strategy of frequently reformulating queries with domain-specific terms.

In studies examining the impact of age on search, differences have been observed between older and younger adults. For instance, \citet{Sanchiz2017seniors} demonstrated that older adults' search queries were less elaborate, significantly incorporating more keywords from the problem statements and adding fewer new keywords. They were less efficient, needing more time to complete search tasks. 
Children also encounter unique challenges in their search, primarily due to their in-development vocabulary and cognitive skills, resulting in misspelling and more intuitive/natural queries. \citet{BILAL2018Children} highlighted that children predominantly use phrase and question-like queries, particularly younger children. This aligns with earlier log analysis research by \citet{Torres2010children}, which suggests that children's queries are longer, more natural, and less effective than those of adults.

The concept of ``persona'' was first introduced by \citet{cooper2003face} as a design tool and has since been useful in establishing a user-centric focus. A persona describes user characteristics and goals, ideally based on an understanding of target users. While widely used in user interaction design, limited research uses personas to help understand, guide, and evaluate the design and development of \ac{IR} systems \cite[e.g.][]{li-etal-2016-persona}.

\subsection{Test Collections and Query Variants}
\ac{IR} system effectiveness is commonly measured using offline test collections, which typically represent each information need with one query. This \emph{single query assumption} underpins the collection of relevance judgments {\cite{v02clef}} and has contributed to a consistent and reusable evaluation environment. However, in recent years, there is evidence of the importance of query variants on system evaluation \cite{DBLP:conf/sigir/AlaofiGMSSSSW22,Culpepper,penha_evaluating_2022, Pera2023Children, DBLP:conf/sigir/BaileyMST17}.

Given the recent promising outcomes from using \ac{LLM}s in relevance labelling \cite{thomas2023large,DBLP:conf/ictir/FaggioliDCDHHKK23} and user query simulation \cite{Alaofi23GptVariants}, we are prompted to question this assumption. This leads us to consider whether we can move beyond assuming a single query will represent a large and diverse range of users, whose information-seeking behaviour varies widely.  

Studies show that users issue a substantial number of query variants despite having the same information need \cite{DBLP:conf/sigir/BaileyMST16, Mackenzie}. The utility of variants in identifying relevant documents dates back several decades~{\cite{sb77cambridge}} and was explored in previous TREC tracks ~{\cite{bw99trec}}. Their effect in constructing document pools is demonstrated to be comparable to that of systems \cite{Moffat}. Recently, \citet{Alaofi23GptVariants} used an \ac{LLM} to simulate human queries for building similar document pools.
Research on variants faces challenges. Methods like crowd-sourcing \cite{Mackenzie,DBLP:conf/sigir/BaileyMST16} and click graphs \cite{zhang_generic_2019} have been used to sample variants from populations and datasets. However, these methods have limitations: crowdsourcing is costly to scale, and click graphs are noisy and lack user properties and context. Given \ac{LLM} capabilities in simulation, this work explores \ac{LLM} ability to generate variants.  

Beyond generating relevance labels in \ac{IR}, several studies have used \acp{LLM} to create synthetic data, including queries \cite{Penha2023Ret, breuer2024data, ran2025two, dai2022promptagator, InParsLuiz}, documents \cite{ Askari2023SynthDoc}, and relevance explanations \cite{Fernando2023RelExplanation}, aiming to build more effective or less biased retrieval models. The generated data is optimised to maximise scores but does not necessarily represent real users. The question of how we can use \ac{LLM} capabilities to simulate the queries of various user demographics and assess their value for system evaluation is under-explored.

Drawing on the literature about search engine user characteristics, and inspired by the use of personas as a user-centered design tool, this study uses an \ac{LLM} to integrate the human element into evaluation practices, offering a diverse representation of users and thus following a more `pessimistic' approach to system evaluation \cite{Diaz2024Pessimistic}, moving beyond average utility to examine a range of query variants of varying quality to better understand system behaviour.

\section{Experiment Design}
\label{exp}

\subsection{Test Collection Selection}
We sourced our \textit{seed queries}, the basis queries for generating query variants, from the passage retrieval tasks in \ac{DL21}~\cite{DBLP:conf/trec/Craswell0YCL21} and \ac{DL22}~\cite{DBLP:conf/trec/Craswell0YCLVS22}. This track is used for benchmarking ad hoc passage retrieval methods. Both years use the expanded MS MARCO dataset (v2), which contains 138 million passages \cite{DBLP:conf/nips/NguyenRSGTMD16}, about 16 times larger than v1.

The DL21 track, due to the increased corpus size, 
included a larger number of relevant passages, leading to concerns about score saturation and re-usability \cite{DBLP:conf/sigir/VoorheesCL22,DBLP:conf/trec/Craswell0YCL21}. To address these issues in DL22, the organizers implemented several strategies, including the selection of more challenging topics, specifically, those that did not contribute to the development of the MS MARCO corpus. The DL21 and DL22 tracks include 53 and 76 queries, respectively, with relevance judgments on a four-point scale. The passage corpus for our experiments is obtained through the ir\_datasets tool \cite{macavaney2021simplified}.

\subsection{Query Variant Generation}
We use three methods for generating query \textit{variants} from seed queries. They follow the same approach but are differentiated by the properties of their transformation profiles.
\begin{itemize}[leftmargin=*,topsep=5pt]
    \item \textbf{Persona-based}. Six personas are chosen with varying properties of age, education, first language, and search modality preference (e.g., mobile or voice). Each persona is described by background information and some search-influencing factors.
    \item \textbf{User group-based}. Eight different user groups are chosen to reflect groups of users sharing one single property (e.g., age). As with personas, we use age, education, first language, and search modality preference. In contrast to the use of personas, this approach could be used to associate metric variability with those properties in isolation.
    \item \textbf{Textual transformation-based}. This category does not relate directly to user profiles but instead stems from a taxonomy of variants developed by \citet{penha_evaluating_2022}. Seed queries undergo four distinct textual transformations: paraphrasing, which involves the replacement of words; order change, where the sequence of words is shuffled; naturality, where queries are transformed into either keyword-based or natural language forms; and misspelling, where seed words are misspelled.
\end{itemize}
    
 These methods represent varying levels of abstraction over users or queries, enabling flexible evaluation depending on available user information.  Persona-based variants may suit scenarios with detailed user information, user group abstractions could be applied when only partial traits are known, and textual transformations offer a simple option when no user data is available. 

\begin{figure}[t]
    \centering
    \setlength{\abovecaptionskip}{1.5pt}
    \setlength{\belowcaptionskip}{1.5pt}
    
    \includegraphics[width=\columnwidth]{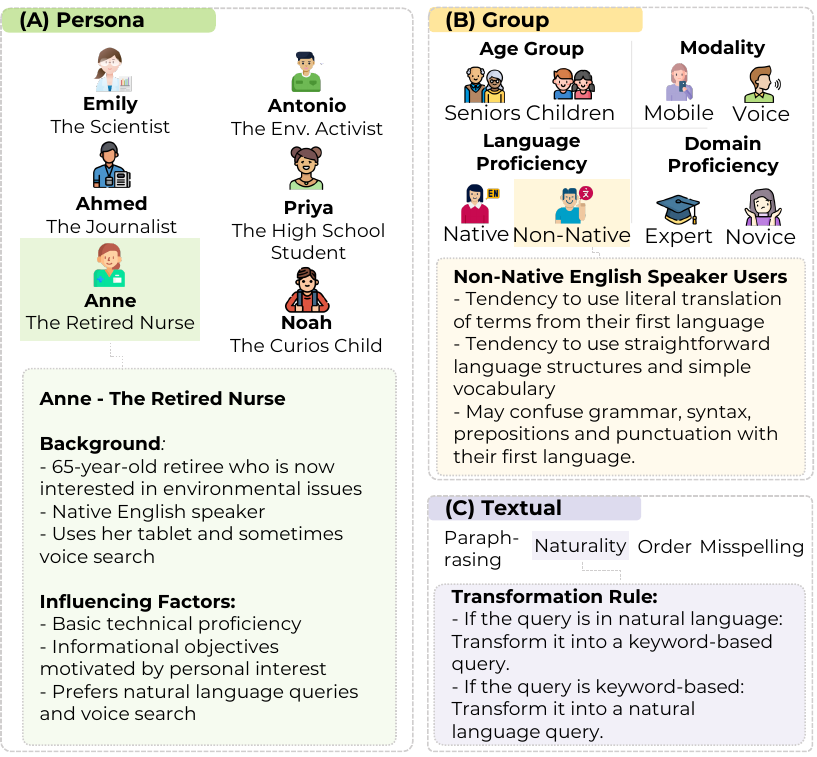}
    \mycaption{Transformation profiles and example descriptions.}
    \label{fig-profile-examples}
\end{figure}

\noindent Figure  \ref{fig-profile-examples} presents all of the profiles associated with each method, together with one complete example profile description for each. Other profile descriptions are publicly available to foster community research and support reproducibility.

\begin{figure}[t]
    \centering
    \setlength{\abovecaptionskip}{1.5pt}
    \setlength{\belowcaptionskip}{1.5pt}
    
    \includegraphics[width=\columnwidth]{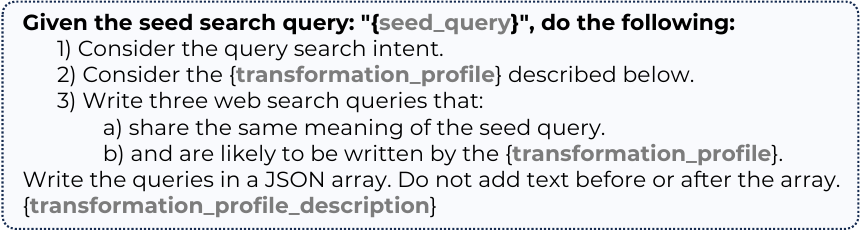}
    \mycaption{The prompt template used to generate variants.}
    \label{fig-prompt}
\end{figure}

In all methods, seed queries together with the transformation profiles are fed into an \ac{LLM}, specifically the GPT-4 model, with the temperature parameter set to one. Initial experiments conducted at lower temperatures yielded more deterministic results, but these were marked by substantially less variation, and with models often getting stuck in repetitive word sequences. Initial human assessments of the temperature set to one yielded positive results, thus we continued with this setting. The model is prompted to generate three variants for each seed and transformation profile combination. The template used to create all prompts is given in Figure \ref{fig-prompt}.

In addition to the profile-based variants, we generate \textit{neutral query variants}, which are generated by GPT-4 without any transformation profiles. These variants help to isolate the impact of GPT-4's capabilities in generating variants, and separate this from the influence of the transformation profiles. For this purpose, we use a similar prompt as shown in Figure \ref{fig-prompt}, but remove the instructions specific to the transformation profile (2 and 3b). The total number of variants generated by each method is given in Table \ref{tbl-variant_stats}. 

\begin{table}[t]
\centering
\small
\setlength{\tabcolsep}{3pt} 
\begin{tabular}{lrrrrr}
\toprule
\multicolumn{1}{c}{\multirow{2}{*}{\textbf{Dataset}}} & \multicolumn{1}{c}{\multirow{2}{*}{\textbf{Seed}}} & \multicolumn{4}{c}{\textbf{Query Variant Sets}} \\
\cmidrule{3-6}

                 &               & \textbf{Persona} & \textbf{Group} & \textbf{Textual} & \textbf{Neutral} \\
\midrule
DL21 & 53 & 954 & 1272 & 636 & 159 \\
DL22 & 76 & 1368 & 1824 & 912 & 228 \\
\bottomrule
\end{tabular}
\mycaption{Number of queries (seeds) and variants as generated by the transformation methods and a neutral setting.
}
\label{tbl-variant_stats}
\end{table}

\subsection{Human Assessment of Query Variants}
\label{QueryVariantAssessment}
When assessing query variants, our objective is to quantify three aspects: (1) \textit{Lexical similarity} to seed queries, where lower levels are preferable; 2) \textit{Semantic similarity}, ensuring that variants retain the same meaning as their seed queries; and 3) \textit{Profile alignment}, whereby variants represent their respective profiles.

To evaluate lexical similarity (1), we use the Jaccard Index, which calculates the ratio of common words to the total words in both seed queries and their variants. To improve accuracy, we first convert all queries to lowercase, remove punctuation, and truncate words using the Porter stemmer~\cite{porter1980algorithm}. Other metrics, such as BLEU and ROUGE, were found to lead to congruent conclusions; for brevity, we therefore report just the Jaccard Index.

To evaluate semantic similarity and profile alignment (2 and 3), we conducted a human assessment on a randomly selected sample of 10\% of the variants created using each transformation profile. 

We recruited annotators from Prolific,\footnote{https://www.prolific.com/}, choosing those whose native language is English and who had an acceptance rate of 98\% or higher. Each annotator was presented with pairs of seed queries and their corresponding variants. The pairs were divided into small batches. Two annotators were assigned to each batch. The tasks for the annotators were twofold: 1) to assess the semantic similarity of each variant to its seed query, and 2) to identify if the variant matches its corresponding profile. A pair is deemed similar if both annotators independently agree on this assessment. The accuracy of semantic similarity is then calculated by dividing the number of similar instances by the total sample size. Further details about the profile alignment task are described in the following section. The \textit{Order} change and \textit{Misspelling} profiles were not assessed for semantic similarity as their transformations are at the lexical level.    
To ensure accuracy and prevent fatigue, the batches were spread over multiple days, with each batch having a maximum of two transformation profiles. Annotators were compensated at a rate of \$15 per hour, based on an estimate of the time needed to complete each from pilot studies.\footnote{This protocol was approved by RMIT University Human Research Ethics Committee.}

For quality assurance purposes, gold standard questions were integrated into the assessment process, with one such question included for every 10 query pairs. These gold questions were handcrafted to feature seed queries paired with queries that share words but differ in meaning, e.g., \textit{``what foods should you stay away from if you have asthma''},\textit{``what types of food is good for fat loss?''}. While all contributed annotations were compensated, only those from annotators who correctly answered all gold questions were ultimately used for analysis. 

For profile alignment, we also aim to compute three features of each transformation profile: 1) variant length; 2) readability, as measured by the Flesch-Kincaid Grade Level Score \cite{kincaid1975derivation}); and 3) lexical diversity, to measure the variety of unique words in profile variants, calculated as the ratio of distinct words to the total word count. To prevent artificially inflating the ratio, misspelled words in the \textit{Child} queries are corrected. We would expect, for example, profiles representing children to demonstrate higher readability and smaller vocabulary diversity compared to those representing adults. These features are reported exclusively for user-specific (persona and user group-based) profiles, as they appear relevant in that context. The Mann-Whitney U test is used to assess the pairwise differences in mean scores of these features across all profiles, with a significance level \(\alpha\) set at 0.05.

\subsubsection{Human Assessment of Profile Alignment}
Assessment was structured slightly differently depending on the method used for generating query variants -- although, in all cases, accuracy was calculated as the number of correct instances divided by the total sample size:

\myparagraph{Persona-based}
For each variant, annotators were asked to identify which persona was more likely to have generated the query variant. Annotators were presented with three options: the correct persona; a candidate persona randomly chosen from the remaining five; and, an option indicating that both personas are ``equally likely''. The variants of each persona are ultimately paired with all other personas as candidates.

The objective was not to measure how distinguishable a persona was but rather to measure if the variants appeared to have been written by the persona. A choice of ``equally likely'' was considered the same as the correct persona. The choice of a persona was deemed correct if both annotators agree on the same persona which matches the correct persona. Any other scenario is considered incorrect. 

\myparagraph{User group-based} For each sampled variant, annotators had three options: the correct user group, their opposite group, and ``equally likely''. For instance, with non-native English speakers, the options were ``native'', ``non-native'', and ``equally likely''. 

\myparagraph{Textual transformation-based}
For \textit{Paraphrasing} and \textit{Naturality} transformation profiles, annotators were presented with a description of the transformation profile. They were then asked to assess whether the variants were transformed in accordance with the given description. For \textit{Order} change and \textit{Misspelling}, the validation was conducted automatically across the entire set of variants, unlike the 10\% sample used for other transformation profiles. A variant is considered to have a different order if the arrangement of words differs from the seed, i.e., the variant is not equal to the seed but a sorted list of the seed's words should match a sorted list of its variant's words for the transformation to be considered valid. For misspellings, we utilised the Bing Spell Check API.\footnote{https://www.microsoft.com/en-us/bing/apis/bing-spell-check-api} A transformation was deemed valid if the variant introduced one or more misspelled words that, when correctly spelled, are part of the seed query.

\subsection{Running Seeds Variants on IR Systems}
Experimental comparisons are conducted across 15 representative \ac{IR} systems: three lexical models (TF-IDF, DLH and BM25) both with and without RM3 and BO1 query expansions; four pre-trained neural re-rankers (BERT \cite{Devlin2019BERT}, ColBERT~\cite{colbert2020Khattab}, ELECTRA \cite{Clark2020ELECTRA}, and monoT5 \cite{Pradeep2021T5}); one neural-augmented index (Doc2Query\cite{Rodrigo2019doc2query}); and one dense model (ANCE~\cite{xiong2020approximate}). This selection enables meaningful comparisons, where practitioners may consider evaluating and comparing both traditional and more advanced, less efficient systems.

Retrieval was performed using Pyterrier \cite{pyterrier2020ictir}, except for \linebreak Doc2Query for which Pyserini \cite{Lin_etal_SIGIR2021_Pyserini} was used over a pre-built corpus with doc2query-T5 expansions.

\subsection{System Evaluation}
\label{sec-sys-eval}
Evaluation is based on NDCG@10, the official metric for \ac{DL21} and \ac{DL22}. Systems were evaluated across all variant sets: seed, persona-based, user group-based, and textual transformation sets. As anticipated, we encountered a substantial portion of missing relevance judgments in the top ten results retrieved in response to query variant sets. On average, at a cutoff of 10, 41\% and 48\% of relevance judgments are missing in \ac{DL21} and \ac{DL22}, respectively, raising concerns about the evaluation outcomes.

With the promising results of using \acp{LLM} to provide relevance labels \cite{thomas2023large, DBLP:conf/ictir/FaggioliDCDHHKK23}, we used an \ac{LLM} for relevance labelling. Based on extensive comparisons and gullibility tests of various \acp{LLM} as conducted by \citet{Alaofi2024LLMs}, we selected GPT-4o, which demonstrated the highest agreement with human judgments and the greatest robustness against keyword-stuffing gullibility tests given the same test collections we are using in this study.

We generate relevance labels for the top ten passages of all systems in response to all query variant sets. We used the same \textit{Basic} prompt used in \citet{Alaofi2024LLMs} and used the same 4-point scale of DL21 and DL22. We initially provided seed queries to GPT-4o to judge passage relevance, but noticed a measurable bias favouring retrieval using seed queries and thus opted to generate backstories to represent the information needs and use them instead of seed queries to assess passage relevance, following  \citet{DBLP:conf/sigir/BaileyMST16}. Backstories are available for public use.  

We assessed the performance of GPT-4o's relevance labels against the available \ac{NIST} relevance judgments using \ac{MAE}, and evaluated agreement with \ac{NIST} judges using Cohen's $\kappa$ \cite{cohen1960coefficient} and Krippendorf's  $\alpha$ \cite{krippendorff2011computing}. When relevance scores are binarised, scores of 0 and 1 are mapped to 0, while scores of 2 and 3 are mapped to 1. The results are presented in Table~\ref{tbl-gpt-performance}.

\myparagraph{GPT-4o labelling performance}
The performance of GPT-4o on a binary scale ($\kappa$ of 0.50 and 0.46) is comparable to human-to-human agreement levels, reported as 0.52 on the OHSUMED dataset~\cite{Hersh1994OHSUMED} and 0.41 on the TREC-6 ad hoc track~\cite{Cormack1998Efficient}. Similarly, its agreement on a graded scale ($\alpha$ of 0.58 and 0.62) is at the higher end of the observed human-to-human agreement range, which spans from 0.41 to 0.69 as measured on the TREC 2004 Robust Track \cite{gauging-2017-Damessie}.
These results indicate that the moderate level of agreement between GPT-4o and NIST judges falls within the typical range of agreement observed among human assessors, and thus, GPT-4o is used for relevance labelling.

\myparagraph{Analysis metrics} 
To answer if the choice of transformation profile significantly impacts systems ranking (RQ2.1); e.g., is a system ranking for \textit{Emily} different from that for \textit{Ahmed}, we ran a pairwise analysis where we paired profiles that originate from the same transformation method, and with seeds. For example, pairs from the persona-based transformation method would be \textit{Emily-seed}, \textit{Emily-Ahmed}, \textit{Emily-Anne}, etc. For each pair of profiles, we consider the question of whether systems are ranked differently using \textbf{Kendall \(\tau\)} \cite{kendall1938tau}, a correlation measure used to compare rankings under different conditions -- in this case, across profiles.
    
For each profile, we perform a two-way ANOVA (topic and system effects). We compare each pair of profiles -- considering all possible pairs of systems (i.e., 105 pairs from 15 systems) -- according to the following measures \cite{Ferro2022Test,Moffat2012Test}:
    \begin{itemize}[leftmargin=*,topsep=5pt]
        \item \textbf{\ac{AA}}: Number of instances that agree on which system is significantly better across both profiles.
        \item \textbf{\ac{AD}}: Number of instances that disagree on which system is significantly better across both profiles.
        \item \textbf{\ac{MA}} \& \textbf{\ac{MD}}: Similar to \ac{AA} and \ac{AD}, but significance observed in one profile only.
        \item \textbf{\ac{PA}} \& \textbf{\ac{PD}}: Similar to \ac{AA} and \ac{AD}, but significance not observed.
    \end{itemize}

\noindent We use three-way ANOVA to examine the effect of topic, system, profile, and their interactions on how systems treat different profiles (RQ2.2). 
We set \(\alpha\) to 0.05 in all of our analyses, accounting for multiple comparisons among systems with Tukey's HSD \cite{tukey1949test}.

\begin{table}[t]
\centering
\small
\mycaption{\ac {MAE} and pairwise agreement between \ac{NIST} judgments and GPT-4o labels, measured using Cohen's Kappa (\(\kappa\)) on a binary scale and Krippendorf's Alpha (\(\alpha\)) on a 4-point ordinal scale. 
}
\label{tab:revised_metrics}
\begin{tabular}{l @{\hspace{3em}} l r r r r}
\toprule
\multirow{2}{*}{\textbf{Dataset}} & \multirow{2}{*}{\textbf{qrels \#}} 
  & \multicolumn{2}{c}{\textbf{Binary}} 
  & \multicolumn{2}{c}{\textbf{Graded}} \\
  \cmidrule(lr){3-4} \cmidrule(lr){5-6}
  & & \multicolumn{1}{p{0.9cm}}{\raggedleft MAE} 
    & \multicolumn{1}{p{0.9cm}}{\raggedleft \(\kappa\)} 
    & \multicolumn{1}{p{0.9cm}}{\raggedleft MAE} 
    & \multicolumn{1}{p{0.9cm}}{\raggedleft \(\alpha\)} \\
\midrule
DL21 & 5141 & 0.25 & 0.50 & 0.69 & 0.58 \\
DL22 & 8011 & 0.20 & 0.46 & 0.55 & 0.62 \\

\bottomrule
\label{tbl-gpt-performance}
\end{tabular}
\end{table}

\section{Results and Discussion}
The results address two main research questions: assessment of query variants and the impact of variants on system evaluation.

\subsection{Query Variant Assessment}

\begin{figure}[t]
    \centering
    \setlength{\abovecaptionskip}{0pt}
    \setlength{\belowcaptionskip}{0pt}
    \includegraphics[trim={0 0 0 1.1cm},clip,width=\columnwidth]{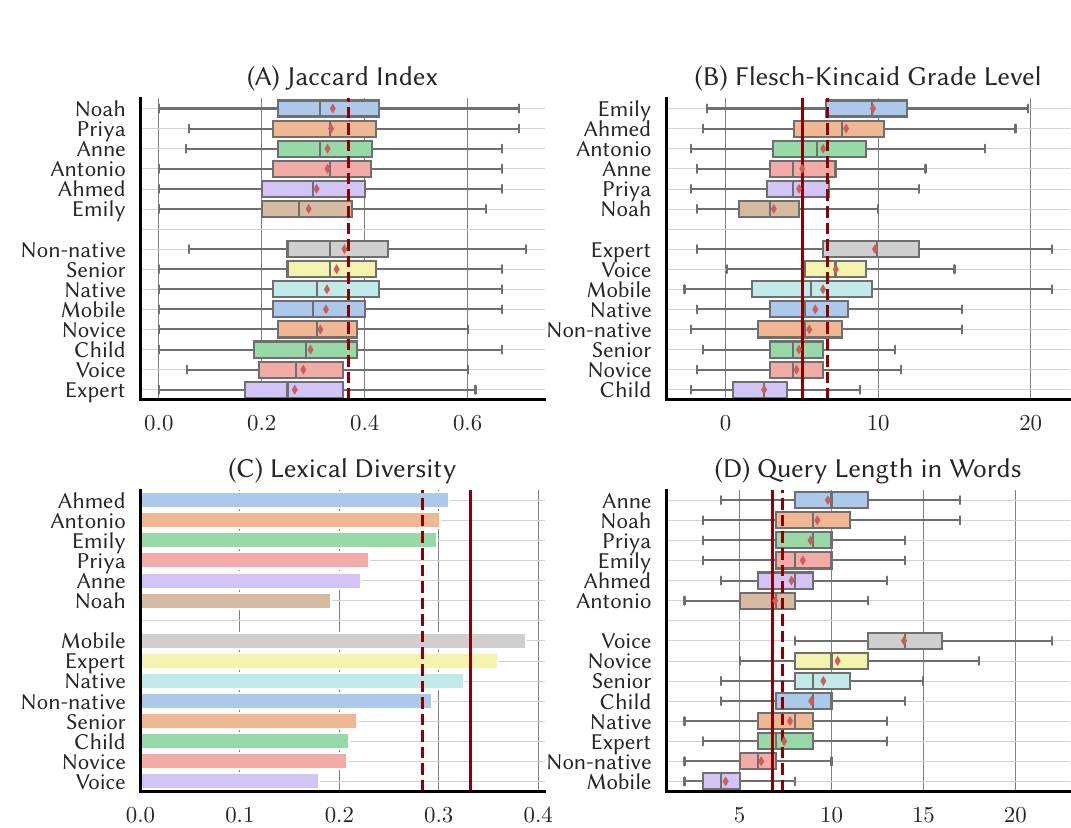}
    \mycaption{
    Distribution of profile variant properties ordered by mean for each transformation method. Vertical lines show the mean value for neutral variants (dashed) and seed queries (solid).
    }
    \label{fig-profile-features}
\end{figure}

\boldheadpara {RQ1.1 Do variants differ lexically from seed queries?}
We quantify the lexical difference between the generated variants and their seed queries using Jaccard Index, where 0 signifies no overlap and 1 maximal overlap (not necessarily identical, due to text pre-processing, see Sec~\ref{QueryVariantAssessment}).
Figure \ref{fig-profile-features}~(A) presents the distribution of Jaccard Index for variants generated by all profiles. The average Index for neutral variants is indicated by a red dashed line. The average Index across all profiles is 0.32. The average is lower for profiles possessing greater domain expertise, i.e., the domain \textit{Expert} and \textit{Emily} the scientist. The impact of using profiles in generating lexically different variants becomes evident when these are compared to neutral variants generated without profiles. Variants from all profiles, except for the \textit{Non-native} user profile, show statistically significantly lower overlaps than the neutral variants.

\boldheadpara {RQ1.2 Are variants semantically similar to seed queries?}
Column one of Table \ref{tbl-profile-alignment} shows the accuracy of semantic similarity based on the human assessment  
of the sampled variants from each transformation profile, paired with their respective seed queries (in total, 611 pairs). Due to the high subjectivity of query semantic similarity assessment \cite{Teevan2010QueryIntent}, we required a consensus between the two annotators for a pair to be considered semantically similar. Disagreements in variant similarity occurred in 9.82\% of the pairs, primarily in cases of highly specific variants, for example \textit{Ahmed} the journalist's query: \textit{``budget for a work trip to Bangkok for a reporter''} for the seed query \textit{``how much money do I need in Bangkok.''} Similarly, ambiguous seed queries could lead to different interpretations by annotators. For example, for the seed query \textit{``Collins The Good to Great''},  \textit{Anne} the retired nurse, and who prefers voice search, and \textit{Noah}, the child, generated the queries \textit{``audio explanation of Jim Collins Good to Great book''} and \textit{``what's the book Good to Great by Collins about?''}. 

Another reason for this disagreement was observed when expert profiles used more scientific terms, e.g., \textit{Emily} the scientist's query \textit{``research studies on magnolia bark extract use for anxiety mitigation''} for the seed \textit{``how much magnolia bark to take for anxiety''}. Country-specific expressions may also contribute to this disparity, for example, \textit{``how to help a \textbf{jammed finger}''}, where a UK-based annotator interpreted the meaning differently than what is commonly understood in American English for a joint injury. Some disagreement is perhaps due to human errors; for example, \textit{``Collins The Good to Great''} and its paraphrased variant \textit{``Collins The Good to Excellent''}, was annotated as similar by one of the annotators. This could be due to a failure to recognise that ``The Good to Great'' is a book title or an assumption that the user may have forgotten the exact title, thereby considering it a valid paraphrase.

Generally, semantic similarity scores are high, substantially exceeding the agreement of both annotators by chance (25\%), with minor shifts observed in some queries of the \textit{Child} and \textit{Expert} profiles. These are primarily attributable to slight oversimplifications or elaborations,  which reflect the nature of these profiles, e.g., \textit{Noah's} query \textit{``can you \textbf{blame} the landlord if a bad guy breaks in and you get hurt''} for the seed query \textit{``are landlords \textbf{liable} if someone breaks in a hurts tenant''}. Nevertheless, the core intent remains intact.

\boldheadpara {RQ1.3 Are variants likely to be generated by their respective profiles?}
Column two of Table \ref{tbl-profile-alignment} presents the accuracy of profile alignment as assessed by the annotators and Figure \ref{fig-profile-features} (B, C and D) show the automatic calculation of query features for profile alignment, as described in Sec~\ref{QueryVariantAssessment}.

It is notable from Table \ref{tbl-profile-alignment} that, for persona-based profiles, \textit{Noah} the child and \textit{Emily} the scientist are most easily distinguishable when paired with other random profiles as set in the experiment. Other profiles also demonstrate a reasonable level of accuracy; however, challenges arise when profiles with similar characteristics are paired. In such cases, the differences between the profiles may not be sufficiently distinct to be fully exhibited in the queries. It is worth noting, though, that all profiles have an alignment accuracy that is substantially higher than the random chance of the agreement of both annotators (44.44\%), and most inaccuracies stem from disagreements, which we consider incorrect—for instance, for \textit{Priya}, only 7.89\% of pairs were misclassified by agreement, while 36.84\% resulted from disagreement.
  
User group-based profiles seem to have a slightly higher accuracy in comparison, which could be due to being paired with opposite profiles. The \textit{Order} change transformation, though simple, does not perform as expected. It adds or removes words in the process, but those changes are minor (mostly in stop words).

In assessing readability, as illustrated in Figure \ref{fig-profile-features} (B), where scores correspond to the school grade level required for understanding, significant variations in means across persona-based profiles are observed. This is true for all pairs except \textit{Anne} and \textit{Priya}. As expected, \textit{Emily}'s queries have the lowest readability, while \textit{Noah}'s have the highest. Other profiles do not show much deviation from the readability means of the seed and neutral queries. Contrary to expectations, the queries produced by non-native English speakers, \textit{Ahmed} and \textit{Antonio}, are not more readable than those by the native speaker, \textit{Anne}.
Similar results are observed for the user group-based profiles: the \textit{Expert} and \textit{Child} profiles produce queries that are significantly less and more readable, respectively, compared to all other profiles, while \textit{Native} and \textit{Non-native} are similarly readable.

\begin{table}[t]
\centering
\small
\begin{tabular}{llrr}
\toprule
& \textbf{Profile} & \textbf{Similarity} & \textbf{\thead{\normalsize Profile Alignment \\ }} \\
\midrule
\multirow{6}{*}{\rotatebox[origin=c]{90}{\textbf{Persona}}} 

& Emily   & 0.97  & 0.87  \\
& Ahmed   & 0.95  & 0.68  \\
& Anne    & 0.92  & 0.70  \\
& Antonio & 1.00  & 0.62  \\
& Priya   & 0.97  & 0.55 \\
& Noah    & 0.92  & 0.73  \\

\midrule
\multirow{8}{*}{\rotatebox[origin=c]{90}{\textbf{Group}}} 

& Child      & 0.67  & 0.74  \\
& Senior     & 0.82  & 0.85  \\
& Non-native & 0.92  & 0.62 \\
& Native     & 0.84  & 0.84  \\
& Novice     & 0.84  & 0.70  \\
& Expert     & 0.72  & 0.79  \\
& Mobile     & 0.92  & 0.95  \\
& Voice      & 0.97  & 1.00  \\

\midrule
\multirow{4}{*}{\rotatebox[origin=c]{90}{\textbf{Textual}}} 

& Naturality   & 0.97  & 0.97  \\
& Paraphrasing & 0.76  & 0.76  \\
& Order        & -     & 0.69 \\
& Misspelling  & -     & 0.80 \\

\bottomrule
\end{tabular}
\mycaption{Human assessment of variant similarity and alignment.}
\label{tbl-profile-alignment}
\end{table}

Figure \ref{fig-profile-features} (C) and (D) show the lexical diversity and the distribution of query length in words for all profiles. As expected, child profiles have lower lexical diversity when compared to most of the other profiles; similarly, \textit{Non-native} is lower than \textit{Native}. \textit{Anne},  \textit{Noah}, and \textit{Priya}, whose profiles prefer natural language queries, have significantly longer queries when compared to other profiles. Not surprisingly, the \textit{Voice} and \textit{Mobile} profiles produced significantly longer and shorter queries, respectively. 

 While this analysis does not verify a direct correspondence between the generated variants and ``real world'' users, a qualitative examination of the query variants reveals some capabilities not only at the linguistic level but also in terms of information-seeking behaviour. For example, for the seed query \textit{``was Friedrich Nietzsche an atheist,''} \textit{Ahmed} the journalist's query aims to find primary sources: \textit{``writings of Nietzsche indicating his atheism.''} 
 Cases of failures exist, too, with a few having distorted meanings or using unusual query openers. 

The results show that we can produce query variants that map their profiles and have properties that align with how those profiles are defined. While one might ask if the variants represent ``real'' users, this is a common concern about many evaluation campaign tasks: in TREC, queries are often sourced from a limited and non-representative group of individuals, with typically no validation to assess how realistically they simulate a broader population of searchers. The method we present here builds on prior research in search behaviour to capture common and meaningful query formulation patterns, and explicitly tests the validity of the generated variants. 
This goes beyond how traditional test collections are built -- including those that capture human variants -- where diversity among participants is hard to obtain and only automated or semi-automated methods are used to assess the quality of queries produced by crowdworkers \cite{DBLP:conf/sigir/BaileyMST16, Mackenzie}. 
Large-scale query log sampling may reflect user diversity but has many limitations. Searchers are anonymous, have varied interests and unclear intent, making it difficult to control for demographics and topics. 

\subsection{System Evaluation}
\begin{figure*}
    \centering
    \setlength{\abovecaptionskip}{0pt}
    \setlength{\belowcaptionskip}{0pt}
    \includegraphics[trim={0 0 0 0.2cm},clip,width=\textwidth]{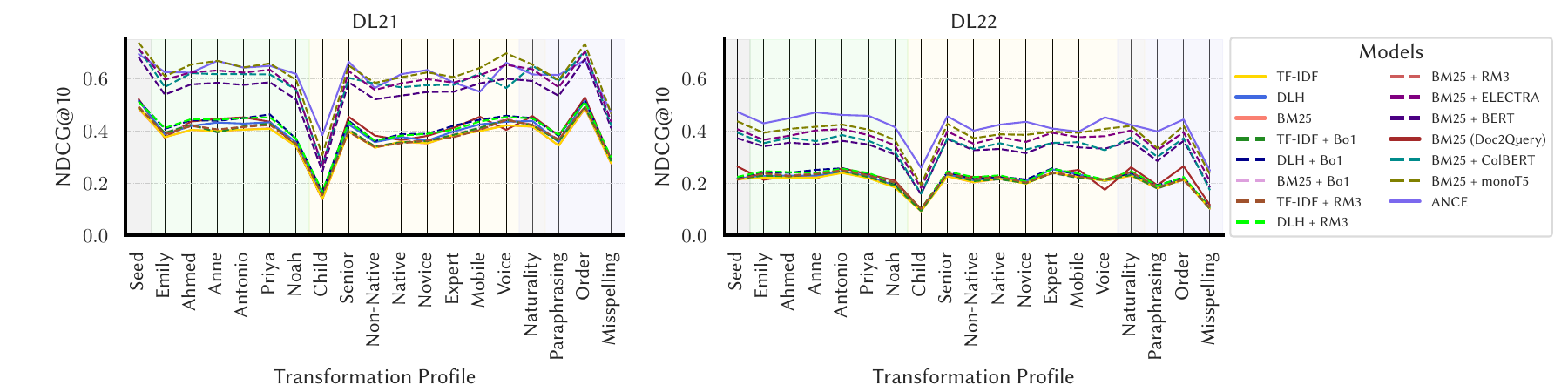}
    
    \mycaption{Selected \ac{IR} systems ranked using NDCG@10 across transformation profiles on \ac{DL21} and \ac{DL22}.}
    \label{fig-system-ranking}
\end{figure*}

    \boldheadpara {RQ2.1 How do query variants impact system ranking?}
    Figure \ref{fig-system-ranking} shows the performance of all systems in response to seed queries and transformation profiles across \ac{DL21} and \ac{DL22}. Performance varies widely depending on the transformation; notably, the \textit{Child} and \textit{Misspelling} profiles lead to the lowest scores across all systems. 
    
   Table \ref{tbl:anova} shows an ANOVA analysis for NDCG@10 of the topic, system, profile factors, and their interactions. The effects are all statistically significant. 
   Profile has a large effect size, implying substantial changes in the overall system performance across profiles. We see this change in Figure \ref{fig-system-ranking}, where a wide range of NDCG@10 values are shown across the profiles. It would appear that there is potential for systems to mitigate such variation in effectiveness. However, the Table shows that the systems used in our test do not perform any such mitigation: the System*Profile interaction shows only a small effect size. This is also visible from Figure \ref{fig-system-ranking}, where system lines cross to a limited extent. Note, however, that the retrieval systems used in the test were not designed with variants in mind. A recent investigation of a novel system that improves the poorly performing variants of a topic appears to show great promise \cite{ran2025two}. Therefore, the small effect size of the profiles on systems observed in this study likely underestimates their value, as they were validated using a set of systems that were not designed to be variant-aware.
   We also observe a large effect size for the Topic*Profile interaction, suggesting that query variants from different profiles will have different levels of influence depending on the topic.

    \definecolor{softblue}{rgb}{0.8, 0.9, 0.95}
\definecolor{strongblue}{rgb}{0.529, 0.808, 0.922}

\begin{table}[t]
\centering
\small
\mycaption{ANOVA effect size $\mathbf{\omega^2_p}$ for nDCG@10. Dark and light shades denote large ($\geq 0.14$) and small ($\leq 0.06$) effects respectively.}
\label{tbl:anova}
\begin{tabular}{lrr}
\toprule
\textbf{Source} & \multicolumn{1}{c}{\textbf{\ac{DL21}}} & \multicolumn{1}{c}{\textbf{\ac{DL22}}} \\
\midrule
Topic & \cellcolor{strongblue}0.7338& \cellcolor{strongblue}0.7741\\
System & \cellcolor{strongblue}0.4439& \cellcolor{strongblue}0.3926\\
Profile & \cellcolor{strongblue}0.3300& \cellcolor{strongblue}0.2189\\
Topic*System & \cellcolor{strongblue}0.3149& \cellcolor{strongblue}0.3470\\
Topic*Profile & \cellcolor{strongblue}0.6713& \cellcolor{strongblue}0.6566\\
System*Profile & \cellcolor{softblue}0.0104& \cellcolor{softblue}0.0156\\
\bottomrule
\end{tabular}
\end{table}

    Comparing the performance of systems on seed queries with other profiles in \ac{DL22}, some profiles generate queries that are more effective (e.g., \textit{Expert} profile with traditional models). The observation supports the human assessment results showing high semantic similarity between variants and seed queries: if variants distort the meanings of the seeds, we would expect a consistent decline in system performance. A different behaviour is, however, observed in \ac{DL21}, where most systems achieve their best performance with seed queries, as measured by NDCG@10. We hypothesize that the observed behaviour may be attributable to a bias in the corpus towards seed queries, a result of integrating relevant passages to the queries from Bing into the corpus, as opposed to \ac{DL22}, where only queries that did not contribute to the corpus were included ~\cite{DBLP:conf/trec/Craswell0YCLVS22}.
    
    To understand how different profiles might alter the conclusions we draw from system effectiveness comparisons. For instance, one might ask if System A outperforms System B in the context of voice search, while exhibiting inferior performance in a mobile query scenario. Table \ref{tbl-kendalls-tau-ndcg} shows the average \(\tau\) of pairwise differences between system rankings across different profiles. A \(\tau\) of 1 indicates complete ranking agreement, -1, ranking is reversed. 
    
    Table \ref{tbl-kendalls-tau-ndcg} also reports the average consistency between pairs of system rankings in the \ac{AA}, \ac{AD}, \ac{MA}  and \ac{MD} groups, as defined in Section~\ref{sec-sys-eval}. The always zero \ac{AD}, complemented by low or zero \ac{MD}, indicates that the profiles do not systematically introduce excessive changes, i.e., leading to completely opposite conclusions regarding relative system effectiveness when using one profile or another. \ac{AA} for around 50\% of the pairs is a good indicator of stability. On the other hand, the \ac{MA} of 1\% to 11\% of the pairs suggests that diverse profiles are able to spotlight diverse significant differences between systems. 
    The average \(\tau\) in most methods given \ac{DL21} and \ac{DL22} is below the threshold of 0.90, which indicates that two test collections (in our case profiles) are not identical in how they rank systems~\cite{Voorhees2001Qrels}.
    This significant but moderate change in the system rankings is consistent with maximum PD being 15\% of the pairs as well as the small size of the System*Profile interaction effect reported in Table \ref{tbl:anova}. Going beyond the metrics, we observe that certain profiles, such as the voice profile in \ac{DL21}, can cause substantial changes in system rankings compared to the seed. Moreover, we can observe that the top-performing system may vary depending on the profile.

    \begin{table}[t]
\centering
\mycaption{Kendall's \(\tau\) correlation of pairwise differences between system rankings and the ratio of \ac{AA}, \ac{AD}, \ac{MA}, \ac{MD} of all profiles of each transformation method
, evaluated using NDCG@10.}
\label{tbl-kendalls-tau-ndcg}
\small
\begin{tabular}{llrrrrrrr}
\toprule
     &           &   \textbf{\(\tau\)} &     \textbf{AA} &    \textbf{AD} &    \textbf{MA}&    \textbf{MD} &    \textbf{PA}&    \textbf{PD}\\
\midrule
& (A) Persona &  0.74&  0.48& 0.00 & 0.01& 0.00 & 0.38& 0.13\\
     DL21& (B) Group &  0.71&  0.48& 0.00 & 0.02& 0.01 & 0.36& 0.14\\
     & (C) Textual &  0.76&  0.48& 0.00 & 0.11& 0.02& 0.30& 0.10\\

\midrule
& (A) Persona &  0.77&  0.50& 0.00 & 0.02& 0.00 & 0.36& 0.12\\
     DL22& (B) Group &  0.69& 0.50& 0.00 & 0.03& 0.00 & 0.32& 0.15\\
     & (C) Textual &  0.69&  0.50& 0.00 & 0.03& 0.00 & 0.32& 0.15\\

\bottomrule
\end{tabular}
\end{table}

\begin{figure}[t]
    \centering
    \setlength{\abovecaptionskip}{0pt}
    \setlength{\belowcaptionskip}{0pt}
    \includegraphics[trim={0 0 0 0.2cm},clip,width=\columnwidth]{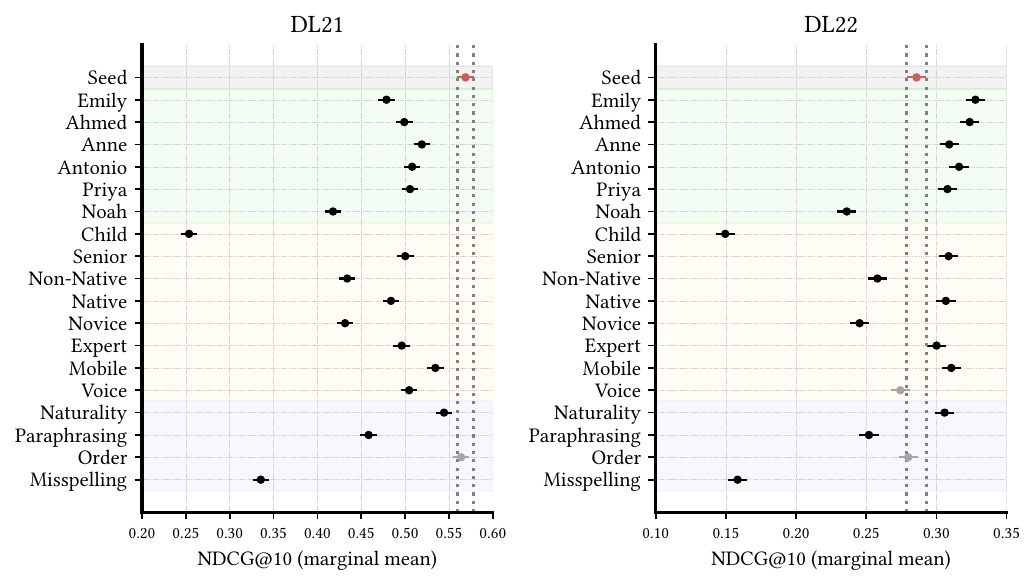}
    \mycaption{Tukey's HSD test on the marginal means of NDCG@10 scores produced by each profile in both \ac{DL21} (left) and \ac{DL22} (right).}
    \label{fig-profile-means}
\end{figure}

    \boldheadpara {RQ2.2 Do systems serve users equally?}
    Figure \ref{fig-profile-means} shows the result of the Tukey's HSD test on the NDCG@10 marginal mean scores of different profiles in \ac{DL21} and \ac{DL22} against the seed queries, averaged across systems, with the error bars representing the confidence intervals around the marginal means.
    The same system bias in favour of seed queries in \ac{DL21} is also present here, aligning with our previous discussions. For this reason, we focus our discussion on \ac{DL22} and refer to \ac{DL21} when needed. Noticeable differences across profiles are observed. When examining persona-based profiles (the green region), we see that \textit{Noah} receives the lowest mean score, substantially lower than all other profiles and significantly lower than the seed queries. Other profiles exhibit different levels of performance, with all showing a significant improvement over the seed queries in \ac{DL22}, which further validates that the generated queries do not drift from the meaning of the seed queries.
    
    Examining user group-based profiles (yellow region), we observe that the \textit{Child} profile has the lowest score, even lower than that of \textit{Noah}. This is most likely because the profile is prompted to generate misspelled words, mimicking a child user. Conversely, the \textit{Senior} profile significantly improves upon the seed mean.
    
    Comparing \textit{Native} and \textit{Non-native} English speaker profiles, we notice that the \textit{Non-native} profile has a significantly lower score than the seed, while the \textit{Native} profile is higher. A similar trend occurs with domain proficiency, where the \textit{Novice} has a significantly lower mean than the seeds while the \textit{Expert} has a significantly higher mean than that of the seeds. 
    
    In modality, \textit{Mobile} queries, characterised by keyword-based searches, demonstrate a significantly higher effectiveness than the seeds, which in \ac{DL21} and \ac{DL22} are more like natural language questions. \textit{Voice} searches do not seem to deviate much from the performance of seed queries, which suggests that longer queries with query openers such as \textit{``Can you tell me''} do not seem to have a significant negative impact on the overall performance of the included systems. 

    The textual-transformation profiles do not map to users but they serve as a reference point in our understanding of effectiveness reductions in user profiles. For example, the change of \textit{Naturality}, which most likely moves the query to a keyword list, shows a similar improvement as was found for \textit{Mobile} queries. Similar to the findings of \citet{penha_evaluating_2022}, \textit{Misspelling} has the highest impact on performance, followed by \textit{Paraphrasing}, \textit{Naturality}, and \textit{Order}. 
    
    These results demonstrate that system performance varies significantly across different user profiles, although the overall impact on system ranking is generally moderate.

\section{Conclusions and Future Work}
We asked the following research questions:

\boldheadpara{RQ1 Are the \ac{LLM}-generated query variants valid?}
Results demonstrate the \ac{LLM}'s ability to generate lexically different, but semantically similar query variants that align with required profiles. This is demonstrated through human assessments, and calculated features such as query length and readability. 

\boldheadpara{RQ2 Do the generated query variants change the outcome of system evaluations?}
Results indicate moderate effects on system ranking, and consistently highlight inequalities among user profiles. 
Notably, profiles such as \textit{Child}, \textit{Non-native}, and \textit{Novice} users experience significantly lower performance.

A limitation of our study is that 
we make a broad assumption by standardising relevance, a practice that is common in the community but not ideal, especially in contexts like ours where we distinguish users through diversified queries. What would be relevant to an \textit{Expert} may not be to a \textit{Novice}. Diversifying the evaluation with relevance labels originating from their issuing profiles would be an interesting avenue for future work.

Also, the studied profiles are not exhaustive; they are hypothetical examples based on properties commonly found interesting in the IR literature. They demonstrate the utility of the methodology presented here, given our limited selection of systems. Certain search engines may require evaluating different competing systems against different user profiles, potentially leading to new conclusions. Future studies can draw on domain-specific needs or taxonomies for user profiles.

The profile alignment is also subject to our profile description, which captures the literature but may not necessarily reflect actual users. This is also impacted by the human assessors' perception of those profiles
(e.g., how an assessor perceives a non-native speaker), despite the provision of detailed instructions. We acknowledge that our validation does not validate that the LLM is generating queries that reflect the user profiles, but rather whether human assessors agree that they do. Although challenging, better validation methods could be explored in future work -- for example, involving intended users directly in the validation, either through annotation or by comparing the generated queries to their actual queries.

Simulating users for evaluating systems, as conducted in this study, may be low-risk, but using such methods to model search behaviours -- such as what users search for or engage with -- can be ethically problematic and may reinforce harmful stereotypes.

The proposed method allows the observation of performance variability and therefore has the potential to enhance the experience of -- or mitigate biases against -- specific groups.  A test collection of this kind may no longer be just a tool to rank systems; it might also help reduce disadvantages for those who struggle to search.

\begin{acks}
  Marwah Alaofi is supported by a scholarship from Taibah University, Saudi Arabia. 
  This research was also supported in part by the CAMEO PRIN 2022 Project, funded by the Italian Ministry of Education and Research (2022ZLL7MW) 
  and the  \grantsponsor{ARC}{Australian Research
  Council}
{https://www.arc.gov.au/} 
  (\grantnum{ARC}{DP190101113}).
  We thank the anonymous reviewers for their helpful feedback.
\end{acks}

\bibliographystyle{ACM-Reference-Format}
\balance
\bibliography{references}
\end{document}